\documentclass[12pt]{article}
\pdfoutput=1

\newenvironment{wileykeywords}{\textsf{Keywords:}\hspace{\stretch{1}}}{\hspace{\stretch{1}}\rule{1ex}{1ex}}

\usepackage[numbers,super,comma,sort&compress]{natbib}
\usepackage{graphicx,color}
\usepackage{amsmath,amsfonts,amssymb}
\usepackage{dcolumn}
\usepackage{lmodern}
\usepackage{comment}
\usepackage[normalem]{ulem}
\usepackage{cancel}

\definecolor{background-color}{gray}{0.98}

\newcommand{\fig}{Fig.}

\newcommand{\figref}[1]{\fig~\ref{#1}}

\renewcommand{\eqref}[1]{equation~(\ref{#1})}
\newcommand{\Eqref}[1]{Equation~(\ref{#1})}

\newcommand{\new}[1]{ #1 }

\newcommand{\bfx}{{\bf x}}

\newcommand{\bfH}{{\bf H}}
\newcommand{\bfB}{{\bf B}}
\newcommand{\bfK}{{\bf K}}
\newcommand{\bfI}{{\bf I}}
\newcommand{\bfP}{{\bf P}}
\newcommand{\bfS}{{\bf S}}

\newcommand{\bfX}{{\bf X}}

\newcommand{\bomega}{{\mbox{\boldmath$\omega$}}}
\newcommand{\bgamma}{{\mbox{\boldmath$\gamma$}}}
\newcommand{\bbiggamma}{{\mbox{\boldmath$\Gamma$}}}
\newcommand{\blambda}{{\mbox{\boldmath$\lambda$}}}
\newcommand{\bbiglambda}{{\mbox{\boldmath$\Lambda$}}}
\newcommand{\boldeta}{{\mbox{\boldmath$\eta$}}}

\title{Instanton rate constant calculations using interpolated potential energy surfaces in non-redundant, rotationally and translationally invariant coordinates.
}
\author{Sean R. McConnell$^*$, Johannes K\"{a}stner\thanks{Institute for Theoretical Chemistry, University of Stuttgart, Pfaffenwaldring 55, 70569 Stuttgart, Germany, sean.mcconnell@theochem.uni-stuttgart.de}}

\date{}

\begin{document}
\maketitle

\begin{abstract}
A trivial flaw in the utilization of artificial neural networks in 
interpolating chemical potential energy surfaces (PES) whose 
descriptors are Cartesian coordinates is their dependence on simple 
translations and rotations of the molecule under consideration. A 
different set of descriptors can be chosen to circumvent this problem, internuclear distances, inverse internuclear distances or z-matrix coordinates are 
three such descriptors. The objective is to 
use an interpolated PES in instanton rate constant calculations, hence 
information on the energy, gradient and Hessian is required at 
coordinates in the vicinity of the tunneling path. Instanton theory relies on
smoothly fitted Hessians, therefore we use energy, gradients and Hessians in
the training procedure. A major challenge is presented in the proper back-transformation of the 
output gradients and Hessians from internal coordinates to Cartesian 
coordinates. We perform comparisons between our method, a previous approach 
and on-the-fly rate constant calcuations on the hydrogen abstraction from 
methanol and on the hydrogen addition to isocyanic acid.
\end{abstract}

\begin{wileykeywords}
Machine learning, neural networks, atom tunneling, reaction rate, instanton theory
\end{wileykeywords}
\clearpage

  \makeatletter
  \renewcommand\@biblabel[1]{#1.}
  \makeatother

\bibliographystyle{apsrev}

\renewcommand{\baselinestretch}{1.5}
\normalsize

\clearpage

%%%%%%%%%%%%%%%%%%%%%%%%%%%%%%%%%%%%%%%%%%%%%%%%%%%%%%%%%%%%%%%%%%%%%%%%
\section{\sffamily \Large Introduction}
%%%%%%%%%%%%%%%%%%%%%%%%%%%%%%%%%%%%%%%%%%%%%%%%%%%%%%%%%%%%%%%%%%%%%%%%

In spite of neural networks gaining ever more attention for their 
ability to interpolate potential energy surfaces (PES) at levels of accuracy approaching 
the highest levels of theory,
\cite{Blank1995,Tafeit1996,Brown1996,No1997,Gassner1998,Prudente1998,Prudente1998a,
Munoz1998,Hobday1999,Cho2002,Rocha2003,Bittencourt2004,Witkoskie2005,Manzhos2006,
Manzhos2006a,Manzhos2006b,Agrawal2006,Doughan2006,Malshe2007,Le2009,Pukrittayakamee2009,
Handley2010,Chen2013,Yuan2016,Schuett2017,Schuett2017a,Lubbers2018,Cooper2018,Laude2018} simple improvements can be 
made to enhance their robustness and range of applications, especially 
for the purpose of instanton rate calculations, an area where little 
work has been conducted to date, though with some exceptions.\cite{Cooper2018,Laude2018} 
External degrees of freedom, i.e. the rotation 
or translation of a molecule do nothing to alter the internal energy of 
the molecule, neither does the exchange of the same atoms within a 
molecule. Neural networks may be trained using a variety of 
different descriptors, for example Cartesian coordinates\new{,
  internuclear distances, z-matrix coordinates} or 
projections 
on to normal mode displacements.\cite{Cooper2018} Though this 
approach produces highly accurate results, it is more 
cumbersome to account for geometrical symmetries.
Ideally, neural networks would be constructed or adapted to 
account for these symmetries, thus allowing for more user-friendly 
interfacing with quantum chemistry programs. In this contribution we 
briefly introduce instanton theory in section \ref{sec:instanton_theory} and thereafter 
focus on how to account for the geometrical symmetries when 
using neural networks in predicting reaction rates using instanton theory. 
Like-atom exchange symmetry may, in specific cases, be 
treated explicitly, for example using symmetry adapted 
internals,\cite{Lorenz2004,Behler2007} but is in general better 
treated using atomistic neural networks.\cite{beh07,Behler2011}

Because we will use the interpolated PES to determine rate 
constants in instanton theory, information on the gradient and 
the Hessian of the surface is also required. To train the neural 
network, we use data obtained from \emph{ab initio} methods. The 
coordinates, gradients and Hessians of the training data are then 
converted into the desired internal coordinate system. A test set is 
also produced using the same approach. Our program 
offers three different input descriptors: internuclear distances, inverse 
internuclear distances and z-matrix coordinates. Usage of internuclear distance based 
coordinates as descriptors is well established in neural networks for 
small molecules,\cite{Malshe2007,Malshe2009,Agrawal2009,Prudente1998,Bittencourt2004,Le2010} 
while z-matrix coordinates, or some combination of internuclear distances, angles 
and dihedrals, also have some pedigree, particularly in atomic 
chain neural networks.\cite{Tafeit1996,Bholoa2007,Sanville2008}
After the training is complete, 
the back-end program, which will use the trained network, reconverts the 
internal gradients and Hessians back into their Cartesian counterparts.

A number of significant differences exist between neural networks 
trained using the descriptors mentioned in comparison to those trained 
on Cartesian coordinates. \new{The training can either be done by
  minimising the error in  Cartesian derivatives or in derivatives with respect to internal
  coordinates. The latter is more natural, because the input data are
  converted into internal coordinates. However, the final interpolated
  PES is used in Cartesian coordinates. Therefore, the Cartesian
  derivatives must be as accurate as possible in the end.}
The transformation of gradients and Hessians from Cartesian 
coordinates to internals introduces a certain amount of numerical noise to the training data, or 
conversely, to the back-converted gradients and Hessians from their 
interpolated internal counterparts. This noise comes as a consequence 
of using pseudoinverses in the transformation process \new{which is
necessary because in general the transformation matrix is highly singular, 
but also because usually there are more Cartesian components than
non-redundant internal coordinate components.} The noise can 
be suppressed to acceptably low levels, the details of which we outline 
in section ~\ref{sec:coords}.

The next major difference is the choice of which internal descriptors 
to use. Z-matrix coordinates require the user to determine a build list, 
usage of internuclear distance based descriptors usually means disregarding 
a large number of potential internuclear distances in order to ensure the set 
chosen is non-redundant. A non-redundant set of descriptors is 
desirable though not necessary. The minimum number of descriptors is 
given by $\min{\left(N\left(N-1\right)/2,3N-N_0\right)}$ where $N$ is 
the number of atoms in the molecule and $N_0$ are the number of 
rotational and translational degrees of freedom. Using a larger number of 
descriptors than this minimum will not result in a poorer interpolation 
but will increase the time required to train the neural network.

The last major difference is the minimisation scheme of the residual 
with respect to the weights and biases. The weights and 
biases are exported to an external program that requires the 
gradients and Hessians be given in Cartesian coordinates. It is 
important that the weights and biases are optimised in 
such a manner as to ensure that the back-transformation 
results in an as small a precision loss as possible. We describe a 
method to achieve this by making a few modifications to the 
traditional residual minimisation scheme.

The PESs to be interpolated are taken from the following reactions: hydrogen 
abstraction from methanol and the hydrogen addition to isocyanic acid. 
These reactions are of interest to the astrochemical community\cite{Goumans2011,ham13,Song2016,cup17} 
and are good candidates for interpolation by neural 
networks due to the relatively low number of atoms involved.

The paper is organised as follows. We describe in detail the 
architecture of the neural network used and the procedure of 
transforming the \emph{ab initio} training data to the desired internal 
coordinate. Next we show how we choose a non-redundant set of the 
internal coordinates and then explain the residual minimisation 
procedure. Finally, we export the interpolated surface to DL-FIND\cite{Kaestner2009a} and 
perform the required rate constant calculations using instanton theory 
and compare with on-the-fly calculations.

%%%%%%%%%%%%%%%%%%%%%%%%%%%%%%%%%%%%%%%%%%%%%%%%%%%%%%%%%%%%%%%%%%%%%%%%
\section{\sffamily \Large Theory \label{sec:theory}}
%%%%%%%%%%%%%%%%%%%%%%%%%%%%%%%%%%%%%%%%%%%%%%%%%%%%%%%%%%%%%%%%%%%%%%%%

%%%%%%%%%%%%%%%%%%%%%%%%%%%%%%%%%%%%%%%%%%%%%%%%%%%%%%%%%%%%%%%%%%%%%%%%
\subsection{\sffamily \large  Instanton theory\label{sec:instanton_theory}}
%%%%%%%%%%%%%%%%%%%%%%%%%%%%%%%%%%%%%%%%%%%%%%%%%%%%%%%%%%%%%%%%%%%%%%%%

Before expounding on the particulars of the theory of neural networks, 
it is necessary to explain why we need an interpolator capable of 
predicting energies, gradients and Hessians. Our rate constants are calculated 
using instanton theory, for the finer details thereof we refer the reader 
to the
literature\cite{Langer1967,Langer1969,Miller1975,Coleman1977,Callan1977,Affleck1981,Coleman1988,Messina1995,Althorpe2011,Richardson2015}
and simply reproduce the equation for the rate constant 

\begin{align}\label{eq:rate}
k_\text{inst}=&\sqrt{\frac{S_0}{2\pi\hbar}}\sqrt{\frac{P}{\beta\hbar}}\frac{\prod_{l=N_0+1}^{NP}\sqrt{\lambda_l^\text{RS}}}{\prod_{l=N_0+2}^{NP}\sqrt{\left|\lambda_l^\text{inst}\right|}}\exp{\left(-S_\text{E}/\hbar\right)},
\end{align}
where there are $N$ degrees of freedom, $N_0$ translational 
and rotational degrees of freedom, $P$ discretisation points 
of the Feynman path (images),
$\beta$ is the inverse temperature ($\beta = 1/k_\text{B}T$), $\hbar$ is 
Planck's constant, $S_\text{E}$ is the Euclidean action along 
the path and $S_0$ is the shortened action.  The values $\lambda^\text{inst}$ 
and $\lambda^\text{RS}$ are the eigenvalues of the second derivative 
matrix of the Euclidean action of the instanton and the reactant 
state, respectively, with respect to all coordinates of all images\cite{Rommel2011a,Rommel2011}
\begin{align}\label{eq:s_e}
\bfS''=\frac{\partial^2S_\text{E}}{\partial q_k^a \partial q_l^b}=&\frac{P}{\beta\hbar}\delta_{a,b}\left(2\delta_{k,l}-\delta_{k-1,l}-\delta_{k,l-1}\right)+\frac{\beta\hbar}{P}\delta_{k,l}\frac{\partial^2E}{\partial q_k^a \partial q_l^b},
\end{align}
where $q_k^a$ is the mass-weighted coordinate component $a$ of image $k$. 
It is worth noting that transforming to mass-weighted coordinates is 
a simple step in Cartesian coordinates, hence the reason why DL-FIND 
uses this coordinate system and why modifying it to accept 
Hessians and gradients in any number of different internal coordinate 
systems is a far more strenuous task than modifying the output of the 
neural network to conform with this pre-existing structure.

The last term in \eqref{eq:s_e} contains the second derivative of the 
energy (Hessian) along the instanton path, hence the reason why we seek 
to interpolate the PES up to the second order spatial derivative. The 
gradient of the PES does not appear in either of the above equations, 
but it too is a necessary part of the procedure to locate the instanton. 
The \emph{most probable} instanton path is defined such that 
$\delta S_\text{E}[y] = 0$, in other words, it is a path of stationary Euclidean 
action, localising this path requires the gradient 
of the PES.\cite{Rommel2011a,Rommel2011}

%%%%%%%%%%%%%%%%%%%%%%%%%%%%%%%%%%%%%%%%%%%%%%%%%%%%%%%%%%%%%%%%%%%%%%%%
\subsection{\sffamily \large  Network architecture \label{sec:architecture}}
%%%%%%%%%%%%%%%%%%%%%%%%%%%%%%%%%%%%%%%%%%%%%%%%%%%%%%%%%%%%%%%%%%%%%%%%
The architecture used in the neural network presented here combines two 
multi-layer perceptrons, one with a single hidden layer (1HL) and one 
with two hidden layers (2HL). The outputs from each network's final 
layer connect to a single node representing the target energy for the 
input geometry.

When speaking of the descriptors, we refer to the internal coordinates 
fed to the 2HL network. A different, and twofold larger, set of 
descriptors are sent to the 1HL network. If we let $\bfx$ be an $I$ 
dimensional vector representing the descriptors of the 2HL network, 
then $\mathrm{x_i^{-1}}$ and $\mathrm{x_{i}^2}$ are the $i^\text{th}$ 
and $i+I^\text{th}$ descriptors of the 1HL network respectively. 
When referencing the 1HL network in formulas, roman upright script will be 
used.

The respective formulas for the $j$, $k$, and $\mathrm{p^\text{th}}$ nodes of
the hidden layers $y^{(1)}_j$, $y^{(2)}_k$, $\mathrm{y^{(3)}_p}$ and the energy $\epsilon$
are given by

\begin{eqnarray}
  y^{(1)}_j         &=& f^{(1)} \left(b_j^{(1)} + \sum\limits_{i=1}^I \left(w_{j,i}^{(1)}\cdot x_i\right)\right),\\
  y^{(2)}_k         &=& f^{(2)} \left(b_k^{(2)} + \sum\limits_{j=1}^J \left(w_{k,j}^{(2)}\cdot y^{(1)}_j\right)\right),\\
  \mathrm{y^{(3)}_p}&=& \mathrm{f^{(3)} \left(b_p^{(3)} + \sum\limits_{i=1}^I \left(w_{p,i}^{(3)}\cdot x_i^{-1}\right) +  \sum\limits_{i=I+1}^{2I} \left(w_{p,i}^{(3)}\cdot x_{i-I}^2\right)\right)},\\
  \epsilon           &=& b^{(4)}_1 + \sum\limits_{k=1}^K\left(w_{1,k}^{(4)}\cdot y^{(2)}_k\right) + \mathrm{\sum\limits_{p=1}^P\left(w_{1,p}^{(5)}\cdot y^{(3)}_p\right)} \label{eq:energy}.
\end{eqnarray}
The symbols $w$ and $b$ are the usual weights and biases connecting the 
layers and $f$ are the transfer functions. A useful notation for this type 
of architecture is $I-J-K\to 1\gets\mathrm{P}-2I$. The superscripts on the weights, 
biases and transfer functions represent their location in 
the neural network, whereas the superscripts on the descriptors $\bfx$ are 
powers. As can be seen, the 2HL network has $J$ and $K$ nodes in the 
first and second hidden layer respectively, while the 1HL network has 
$\mathrm{2I}$ descriptor nodes, and $\mathrm{P}$ hidden nodes. The 
transfer functions $f^{(1)}$ and $f^{(2)}$ are chosen as 
$f^{(1)}(z)=f^{(2)}(z)=\tanh(z)$, the transfer function 
$\mathrm{f^{(3)}}$ is chosen as $\mathrm{f^{(3)}(z)=\tanh(z/2)}$.

The gradients and Hessians are determined from \eqref{eq:energy} 
via the chain rule with respect to the descriptors. With this in mind, 
the cost function, or residual, used here accounts for the inaccuracies 
in the energies, gradients, Hessians \emph{and} Hessian eigenvalues and 
is given by

\begin{align}
  R=&\frac{1}{N_\text{E}+N_\text{G}+N_\text{H}+N_\text{L}}\times\nonumber\\
  &\Bigg[A_\text{E}\sum\limits_{e=1}^{N_\text{E}}\left(\epsilon_e-E_e\right)^2+
  A_\text{G}\sum\limits_{g=1}^{N_\text{G}}\left|\bgamma_g-\bbiggamma_g\right|^2+A_\text{H}\sum\limits_{h=1}^{N_\text{H}}\left|\boldeta_h-\bfH_h\right|^2+A_\text{L}\sum\limits_{l=1}^{N_\text{L}}\left|\blambda_l-\bbiglambda_l\right|^2\Bigg],
\end{align}
where lower-case greek symbols represent the interpolated values 
obtained by the neural network and upper-case greek are the reference 
values in the training set. It is assumed that the reference quantities in  
the training and test sets have been provided in Cartesian coordinates.  
The sum of all the elements squared is used as the norm for matrices, equivalent
to vectors.
A set of weights has also been included 
$\left\{A_\text{E},A_\text{G},A_\text{H},A_\text{L}\right\}$ should there be a need to emphasize 
accuracy in one or more parameters against any other. Furthermore, 
energies, gradients and Hessians are usually generated for one common 
molecular configuration, hence $N_\text{E}=N_\text{G}=N_\text{H}=N_\text{L}=N_\text{train}$.

The reason for the unusual choice of network architecture and usage of 
differing powers of the descriptors is 
to explicitly provide to the network some of the geometrical dependencies 
that influence the properties we wish to interpolate, rather than relying 
solely on the weights and biases to intuit these dependencies. 
For instance, the energy of the system exhibits some 
dependence on the inverse of the internuclear distance, this property can 
be more easily learned by the network since the modulus of the inverse of typical internuclear distances 
is less likely to lead to node saturation by the transfer function.  
Furthermore, 
with a view to future implementations, it may also prove useful to 
use bond angles and dihedral angles as the descriptors for one of 
the two networks, and (inverse) internuclear distances for the other such 
that the size of the parameter space of weights and biases remains 
manageable.

At this point, it is necessary to explain a few particularities in the 
definition of the residual when working in internal coordinates. It 
would seem intuitive that if the descriptors have been 
transformed to an internal coordinate system $\bfx\to\tilde{\bfx}$, 
then one should also transform the reference quantities 
$\left\{\bbiggamma,\bfH,\bbiglambda\right\}\to\left\{\tilde{\bbiggamma},\tilde{\bfH},\tilde{\bbiglambda}\right\}$ into this internal coordinate system and 
obtain an internal residual $\tilde{R}$ (a tilde will be used henceforth 
to signify any quantities in the internal coordinate system). 
One would then minimize this residual with 
respect to the weights and biases and, once a convergence criteria has 
been reached, export the optimised weights and biases to the external 
program where the internal quantities 
$\left\{\tilde{\bbiggamma},\tilde{\bfH},\tilde{\bbiglambda}\right\}$ 
are back-transformed into their Cartesian counterparts.

The alternative is of course to leave the reference quantities in their 
Cartesian representation, back-transform the interpolated quantities 
$\left\{\tilde{\bgamma},\tilde{\boldeta},\tilde{\blambda}\right\}\to\left\{\bgamma,\boldeta,\blambda\right\}$ 
and obtain a residual in Cartesian coordinates. 
It is in fact this approach which is adopted here due to the fact that 
the training of the neural network, based on a Cartesian residual, is 
able to actively adapt the weights and biases to counteract 
uncertainties introduced by the coordinate transform, we give an in 
depth explanation of this property in the following section .
%%%%%%%%%%%%%%%%%%%%%%%%%%%%%%%%%%%%%%%%%%%%%%%%%%%%%%%%%%%%%%%%%%%%%%%%
\subsection{\sffamily \large  Coordinate transformation and training of the neural network \label{sec:coords}}
%%%%%%%%%%%%%%%%%%%%%%%%%%%%%%%%%%%%%%%%%%%%%%%%%%%%%%%%%%%%%%%%%%%%%%%%

We use the following algorithm for coordinate, gradient, and Hessian
transformation:\cite{Bakken2002}

\begin{equation}
  \tilde{\bbiggamma}=\bfP\left(\bfB^{\text{T}}\right)^{\dagger}\bbiggamma,
\end{equation}
with 
\begin{equation}
B_{ij}=\frac{\partial \tilde{x}_i}{\partial x_j} \text{ \ \  and \ \ }
\bfP = \bfB\bfB^{\dagger}.
\end{equation}
$\bfB$, Wilson's B matrix, is in general non-square and potentially highly
singular, therefore $\bfP\neq\bfI$. In the reactions tested in section  
\ref{sec:applications}, $\bfP\sim\bfI$ and therefore is set equal to $\bfI$, 
we include it here and in the following formulas for completeness. 
Owing to the nature of $\bfB$, one utilises
pseudoinverses, indicated by $^{\dagger}$, to derive the reference gradients
and Hessians in internal coordinates. The Hessians are transformed as 
\begin{align}\label{eq:transformations}
  \tilde{\bfH}&=\bfP\left(\bfB^{\text{T}}\right)^{\dagger}\left(\bfH-\bfK\right)\bfB^{\dagger}\bfP,\\
  K_{kj} &= \sum_{i}\tilde{\Gamma}_{i}\frac{dB_{ij}}{d x_k}.
\end{align} 
For inverse internuclear distances $\tilde{x}_i=\left|\bfx_a-\bfx_b\right|^{-1}$,
$\bfB$ has the form
\begin{align}
  B_{ij}=&-\tilde{x}_i^3\sum\limits_{n=1}^3\left(x_{3a-3+n}-x_{3b-3+n}\right)\times\left(\delta_{3a-3+n,j}-\delta_{3b-3+n,j}\right),\\
  \frac{\partial B_{ij}}{\partial x_k}=&\frac{3}{\tilde{x}_i}B_{ij}B_{ik}-\tilde{x}_i^3\sum\limits_{n=1}^3\left(\delta_{3a-3+n,k}-\delta_{3b-3+n,k}\right)\times
  \left(\delta_{3a-3+n,j}-\delta_{3b-3+n,j}\right)
\end{align}
where $\bfx_a$ and $\bfx_b$ refer to the Cartesian coordinates of 
atoms $a$ and $b$, respectively. Hence subscript $i$ maps to a unique 
pair of atoms $a$ and $b$.  

The nature of the pseudoinverse means that some accuracy is lost in performing
the transformation, applying the back-transformation may produce results which
are significantly different compared with the original gradient or
Hessian. This phenomenon is governed by the \emph{condition number} $\kappa$
of $\bfB$, a quantity proportional to the product
$\left\Vert\bfB\right\Vert\cdot\left\Vert\bfB^\dagger\right\Vert$.  A high condition number would mean, for instance, that in spite
of obtaining a quite accurate approximation
$\tilde{\bgamma}\sim\tilde{\bbiggamma}$ after training, the magnitude of
uncertainty $\delta\tilde{\bgamma}$ in the internal representation, upon
back-transformation, is magnified $\kappa$-fold in the uncertainty
$\delta\bgamma\sim\kappa\delta\tilde{\bgamma}$ of the Cartesian
representation. In this work, we make use of Tikhonov
regularisation\cite{Tikhonov1995} in creating pseudoinverses.

In the case of an internal residual, this addition of 
uncertainty to the interpolated Cartesian quantities occurs after 
the training phase, there is no way for the neural network to 
actively compensate for this during the training phase. 
If however a Cartesian residual is used, each step of 
the training phase, i.e. the minimum search on the residual 
hypersurface, is actively guiding the weights and biases to correct for 
the error incurred during transformation.

A more insightful form for the residual, where the interpolated, 
internal quantities are transformed back to Cartesian coordinates, has 
the following form

\begin{align}
  R=&\frac{1}{N_\text{E}+N_\text{G}+N_\text{H}+N_\text{L}}\times\Bigg[
  A_\text{E}\sum\limits_{e=1}^{N_\text{E}}\left(\epsilon_e-E_e\right)^2+
  A_\text{G}\sum\limits_{g=1}^{N_\text{G}}\Big|\underbrace{\bfB^T\bfP^T\tilde{\bgamma}_g}_{\bgamma_g}-\bbiggamma_g\Big|^2+\nonumber\\
  &A_\text{H}\sum\limits_{h=1}^{N_\text{H}}\Big|\underbrace{\bfB^T\left(\bfP^T\tilde{\boldeta}_h\bfP^T+\tilde{\bfK}\right)\bfB}_{\boldeta_h}-\bfH_h\Big|^2+
  A_\text{L}\sum\limits_{l=1}^{N_\text{L}}\Big|\underbrace{\text{diag}\left(\bfX^T\boldeta_l\bfX\right)}_{\blambda_l}-\bbiglambda_l\Big|^2\Bigg],\nonumber\\
  \tilde{\bfK}=&(\bfB^T)^\dagger\bfK\bfB^\dagger,\label{eq:residual}
\end{align}

and $\bfX$ is the matrix of eigenvectors of $\boldeta$. From \eqref{eq:residual} 
it is relatively straightforward to construct $\partial R/\partial \omega_i$ 
and further proceed with the optimisation of the residual. The reason 
for the replacement of $\blambda$ in \eqref{eq:residual} is for the sake of the numerical 
stability of the derivative of $R$ with respect to the weights and 
biases. If we stack all the weights and biases in a vector $\bomega$, 
 take the derivative of $\boldeta$ with respect to $\omega_i$ and 
diagonalise, it is clear that both $\boldeta$ and $\partial\boldeta/\partial\omega_i$ 
must share the same set of eigenvectors since the derivative with respect 
to the $i^\text{th}$ weight/bias $\omega_i$ has no dependence on any spatial 
coordinate. This allows one to use $\partial\boldeta/\partial\omega_i$ 
in both the derivative of the third and fourth sums of 
\eqref{eq:residual}

\begin{align}
  \frac{\partial R}{\partial \omega_i}=&\frac{2}{N_\text{E}+N_\text{G}+N_\text{H}+N_\text{L}}\times\Bigg[&\nonumber\\
  A_\text{E}\sum\limits_{e=1}^{N_\text{E}}&\left(\epsilon_e-E_e\right)\frac{\partial \epsilon_e}{\partial \omega_i}+
 A_\text{G}\sum\limits_{g=1}^{N_\text{G}}\left(\bfB^T\bfP^T\tilde{\bgamma}_g-\bbiggamma_g\right)\circ\bfB^T\bfP^T\frac{\partial\tilde{\bgamma}_g}{\partial\omega_i}+\nonumber\\
 A_\text{H}\sum\limits_{h=1}^{N_\text{H}}&\left(\bfB^T\left(\bfP^T\tilde{\boldeta}_h\bfP^T+\tilde{\bfK}\right)\bfB-\bfH_h\right)\circ
  \underbrace{\bfB^T\left(\bfP^T\frac{\partial\tilde{\boldeta}_h}{\partial\omega_i}\bfP^T+\frac{\partial\tilde{\bfK}}{\partial\omega_i}\right)\bfB}_{\frac{\partial\boldeta}{\partial\omega_i}}+&\nonumber\\
 A_\text{L}\sum\limits_{l=1}^{N_\text{L}}&\left(\text{diag}\left(\bfX^T\boldeta_l\bfX\right)-\bbiglambda_l\right)\circ\text{diag}\left(\bfX^T\frac{\partial\boldeta_l}{\partial\omega_i}\bfX\right)\Bigg],\label{eq:drdw}&
\end{align}

where $\circ$ refers to the Hadamard product. 
\Eqref{eq:drdw} is then used to minimise the residual by means of the 
L-BFGS algorithm.\cite{Liu1989} To obtain 
$\partial\blambda_l/\partial\omega_i$ another possibility is
to directly diagonalise $\partial\boldeta_l/\partial\omega_i$ for every 
$i$. However this approach is encumbered by larger numerical instability, 
and increased computational expenditure 
than taking $\partial\blambda_l/\partial\omega_i=\text{diag}\left(\bfX^T\frac{\partial\boldeta_l}{\partial\omega_i}\bfX\right)$.

%%%%%%%%%%%%%%%%%%%%%%%%%%%%%%%%%%%%%%%%%%%%%%%%%%%%%%%%%%%%%%%%%%%%%%%%
\subsection{\sffamily \large  Finding non-redundant descriptors \label{sec:find_nrc}}
%%%%%%%%%%%%%%%%%%%%%%%%%%%%%%%%%%%%%%%%%%%%%%%%%%%%%%%%%%%%%%%%%%%%%%%%

When working with internuclear distances as descriptors, for the purpose of 
finding a non-redundant set 
of coordinates as well as reducing the computational load in the training
phase, one should 
only use $3N-N_0$ of the $N(N-1)/2$ available internuclear distances as descriptors 
in the neural network. For molecules with five or more atoms, there will 
exist some redundancy in the descriptors if all $N(N-1)/2$ internuclear distances are 
used.

The approach we have adopted is firstly to examine the total number $\mathfrak{C}$ 
of non-redundant combinations of internuclear distances. If this number is less than a 
certain threshold $T$ ($10^7$ is a reasonable figure), 
all combinations will be tested. The formula for $\mathfrak{C}$ is

\begin{equation}
\label{eq:combinations}
\mathfrak{C}=\left\{
	\begin{array}{cc}
		^{N\left(N-1\right)/2}C_{3N-N_0};&N>4\\
		1;&N\leq 4
	\end{array}
\right..
\end{equation}

If $\mathfrak{C}>T$ we randomly choose a combination of $3N-N_0$ internuclear distances 
from the available $N(N-1)/2$ a 
total of $T$ times, importantly however, only those combinations which join 
each atom to at least three other atoms are accepted. 
Once the set of internuclear distance combinations is established, the 
next step is to examine the sum of all internuclear distances for 
this particular combination and average this sum over all the 
configurations in the test and training set. Denoting this quantity 
$\bar{S}$, for a particular combination $k$ of internuclear distances, it is 

\begin{equation}
\underset{1\leq k\leq\min\left(C,T\right)}{\bar{S}_k}=\frac{1}{\nu}\sum\limits_{j=1}^{\nu}
\sum_{i=1}^{3N-N_0}r_{ij,k},
\end{equation}

where it is assumed that $N>4$ and $r$ is an internuclear distance and $\nu=N_{train}+N_{test}$. 
The combination producing the smallest $\bar{S}_k$ is that which is then used as
our descriptors because the change in distances, which are short to start
with, can be assumed to have the largest influence on the total energy of a
molecule. Other approaches exist for finding the descriptors, a quasi
non-redundant set of descriptors can be found by choosing more than the
minimum $3N-N_0$ internuclear distances. Another approach might be to build Wilson's $B$ matrix
for each internuclear distance combination $k$ and molecular configuration $j$, yielding a set
of singular values, the variance of which, averaged over $j$, is as small as
possible.  To that end, it is also possible to select the combination $k$ that
minimises the average variance in $\tilde{\bgamma}$ or $\tilde{\boldeta}$. In
section ~\ref{sec:applications} we present results based on the smallest
$\bar{S}_k$ approach as, for the cases tested, this gave fast convergence of
the residual and produced $\bfB$ matrices which are reasonably
non-singular. Moreover, in the cases tested, $\mathfrak{C}$ is maximal for the
hydrogen abstraction from methanol where it is only 54264, hence we test all
possible combinations to find the one producing the smallest $\bar{S}_k$.

%%%%%%%%%%%%%%%%%%%%%%%%%%%%%%%%%%%%%%%%%%%%%%%%%%%%%%%%%%%%%%%%%%%%%%%%
\section{\sffamily \Large Applications \label{sec:applications}}
%%%%%%%%%%%%%%%%%%%%%%%%%%%%%%%%%%%%%%%%%%%%%%%%%%%%%%%%%%%%%%%%%%%%%%%%

We have applied the neural network described in section ~\ref{sec:theory} 
to the hydrogen addition to isocyanic acid H + HCNO $\to$ H$_2$CNO (R1) and to the hydrogen 
abstraction from methanol CH$_3$OH + H $\to$ CH$_2$OH + H$_2$ (R2). For 
both reactions, the descriptors used are inverse distances. For 
R1, a training set and a test set, comprising energies, 
gradients and Hessians for 90 and 92 different geometries respectively, 
was created. We used density functional theory with the 
BHLYP-D3\cite{Becke1993,Lee1988,Grimme2010} functional and the 
def2-TZVP\cite{Weigend2005} basis set for the training and test data, 
the geometries selected were based on instanton calculations for the 
temperature range 285 K to 70 K. The instanton calculations were 
performed on the fly. The network architecture for R1 is 
9-18-18$\to$1$\gets$40-18, and the weight parameters were 
$A_\text{E}=1,A_\text{G}=5,A_\text{H}=1,A_\text{L}=2$ in atomic units.

\begin{figure}
\begin{center}
  \includegraphics[width=0.9\columnwidth]{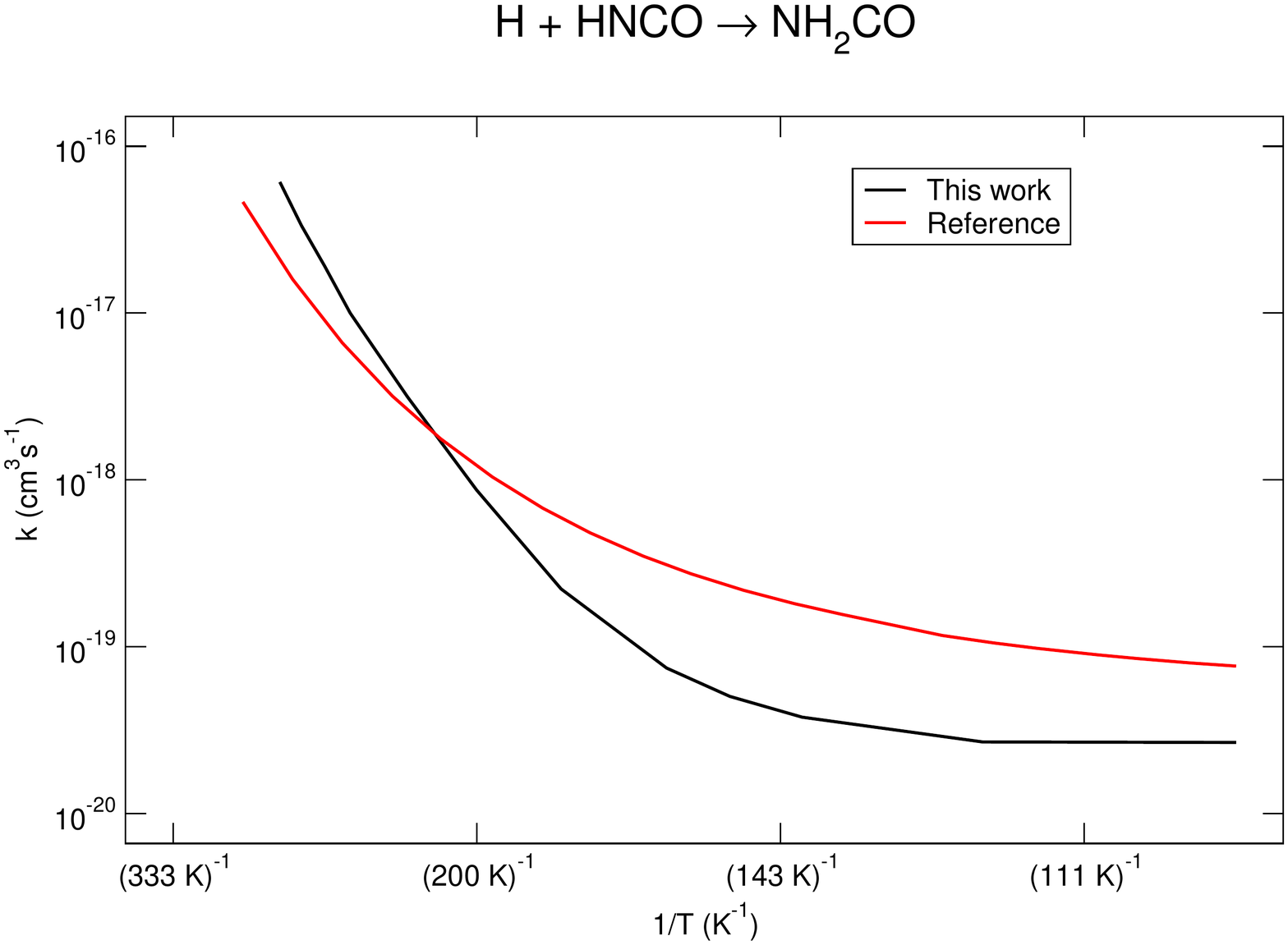}
  \caption{\label{fig:h2nco_rate}Reaction rate constants for the hydrogen 
  addition to isocyanic acid on the neural network PES (black curve), 
  instantons comprised of 77 images were used. Reference curve from 
  \citet{Song2016}.
  }
\end{center}
\end{figure}

The reaction rate constant for R1 is shown in \figref{fig:h2nco_rate}.  The
black curve, representing rate constants calculated on the interpolated neural
network PES is never more than one half of an order of magnitude away from the
rate constants calculated on the fly (red curve).  Since the rate constant
  depends strongly on many details of the potential energy surface along the
  instanton path, a higher interpolation accuracy is difficult to achieve. We
  expect the intrinsic errors of the underlying quantum chemical methods as
  well as the semiclassical approximation inherent to instanton theory to lead
  to errors in a similar range.\cite{Kaestner2014}

The main source of deviation in the rate constants from the neural network for R1 
comes from the fact that $\bfS''$ 
contains more than the expected 6 modes with a non-negligible 
projection on the translation and rotation modes and hence more than the 
expected 6 zero (or nearly zero) eigenvalues. This makes the identification 
of which eigenvalues to exclude in \eqref{eq:rate} a more difficult task. 
It is for this reason 
that we set $A_\text{L}=2$, in order to encourage the neural network to 
emphasize the correct identification of eigenvalues and thereby improve 
the accuracy of the interpolated Hessians.

\begin{figure}
\begin{center}
  \includegraphics[width=0.9\columnwidth]{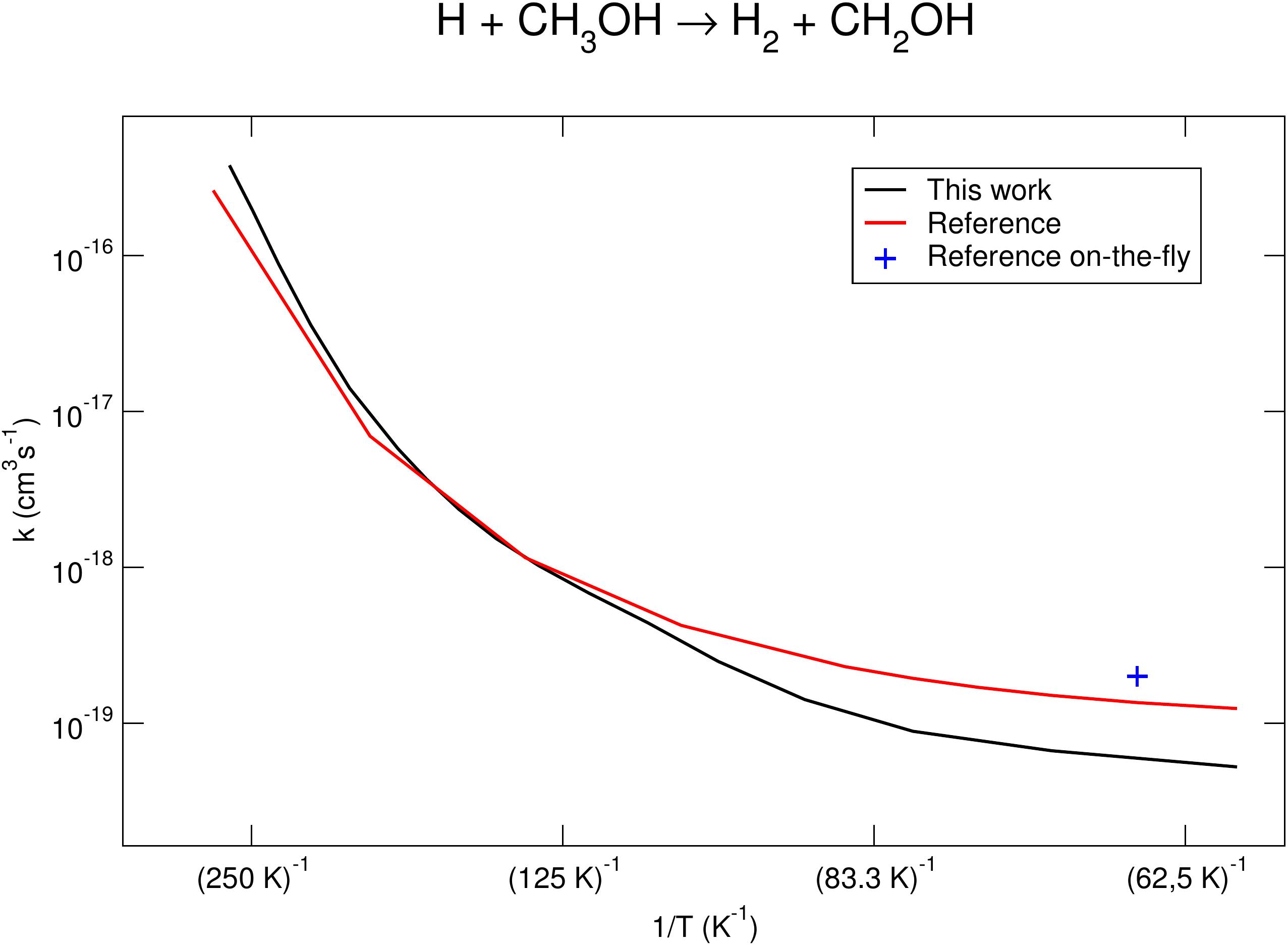}
  \caption{\label{fig:methanol_rate} Reaction rate constants for the hydrogen 
  abstraction from methanol on the neural network PES (black curve), 
  instantons comprised of 100 images were used. Reference curve from
   \citet{Cooper2018}.
  }
\end{center}
\end{figure}

For R2, a training set and a test set was created for 48 and 42 different
geometries, respectively. The chosen target level of theory is
UCCSD(T)-pVTZ-F12, on a restricted Hartree-Fock (RHF) basis, gradients and
Hessians were obtained by finite differences of energies.  These are the same
training and test sets used previously.\cite{Cooper2018} The computational
requirements at this level of theory prohibit extensive on-the-fly instanton
optimisations. The rate constant of a single instanton optimised
  on-the-fly at 65~K is shown in \figref{fig:methanol_rate}. The network
architecture for R2 is 15-30-30$\to$1$\gets$60-30 and the weight parameters
were $A_\text{E}=A_\text{G}=A_\text{H}=1,A_\text{L}=0$ in atomic units. The
reaction rate constants for R2 are shown in \figref{fig:methanol_rate}.

The accuracy for R2, compared with the red curve, for which elongations along normal modes have
been used as descriptors,\cite{Cooper2018} is very favourable. For all temperatures the 
deviation is less than the half an order of magnitude. 
We note that the red curve is the result of an average of 
multiple neural network calculations, while our result required 
only a single network to be trained. More training runs with different starting
weights lead to similar rate constants.

In both reactions there seems to be very little noise in the 
rate constant curve, indicating that the Hessians are also quite noise-free. 
This is, of course, the desired outcome and shows the usefulness of 
\eqref{eq:drdw} in the training procedure, and the tendency of 
Tikhonov regularisation to minimise uncertainties generated as part of 
the pseudo-inversion operation.

%%%%%%%%%%%%%%%%%%%%%%%%%%%%%%%%%%%%%%%%%%%%%%%%%%%%%%%%%%%%%%%%%%%%%%%%
\section{\sffamily \Large Discussion}
%%%%%%%%%%%%%%%%%%%%%%%%%%%%%%%%%%%%%%%%%%%%%%%%%%%%%%%%%%%%%%%%%%%%%%%%
Perhaps the most important aspect of any machine-learning-based approach to
modelling (not simply in quantum chemistry) is to provide high quality
training data. It is not difficult to imagine how error-prone it could be when
attempting to create a training database for a PES if one is required to first
orient and displace the molecule of interest in precisely the form prescribed
by the neural network. The main objective of this contribution is to prove
that a neural network interpolator, capable of accurately predicting energies,
gradients and Hessians, can be created that is independent of trivial
rotations and translations of the molecule being modelled. As the results in
section \ref{sec:applications} have shown, this objective is indeed
achievable.

Another goal for PESs interpolated by neural networks is for the 
set up to be as free of user-input as possible. For the network shown 
here, this is only partially the case, a number of parameters must 
be chosen by the user such as network architecture, layer sizes, 
residual weights $\{A_\text{E},A_\text{G},A_\text{H},A_\text{L}\}$, transfer functions, 
pseudoinversion scheme and internal coordinate system. The system 
sizes used here are small enough that one may make a number of 
reasonable guesses for these parameters, initiate a training run for 
each guess, and simply export the weights and biases for that run which 
gives the smallest residual. This does not, however, transfer easily to 
reactions involving ten or more atoms as the time required for each 
training step increases sharply with system size.

For larger systems, the first computational hurdle is reached in 
finding a non-redundant set of internuclear distances 
(cf. \eqref{eq:combinations}). As 
the system size increases, a smaller fraction of the number of potential 
non-redundant combinations can be tested, increasing the chance 
that more favourable combinations are not utilized, hence increasing 
the number of steps required in the training phase before an acceptably 
small residual is reached, but for systems of less than or 
approximately 10 atoms it is not too problematic to simply test all possible 
non-redundant internuclear distances.

The second computational restriction is the increase in the size of 
the layers as the system size increases. A good rule of thumb for the 
extent of the hidden layers is to have twice or three times as many hidden 
nodes as input nodes, if we apply this to a system containing say, 15 atoms, 
we need to optimise 21685 weights and biases (using the network 
architecture described in \ref{sec:architecture}). Given in tables \ref{tab:comp_costNR} and \ref{tab:comp_costR} is an overview of the difference in computational loads in both the pre-processing phase and the optimization phase when using non-redundant/redundant sets of descriptors, the calculations are performed on an Intel(R) Core(TM) i5-4590 CPU @ 3.30GHz. An additional reaction not 
explored in the applications section, i.e. ring-opening of the cyclopropylcarbonyl radical, 
has been included in the tables for performance comparison purposes.

\begin{table}
	\begin{tabular}{|l|l|p{3cm}|p{4cm}|}
		\hline	
		Reaction&Architecture&Pre-processing time (s/input)&Average optimization time (s/input/step)\\
		\hline	
		R2&15-30-30$\to$1$\gets$60-30&$6.73\times10^{-2}$&$9.16\times10^{-2}$\\
		\hline	
		Cyclopropylcarbonyl&27-30-30$\to$1$\gets$60-54&$9.00\times10^{-1}$&$6.56\times10^{-1}$\\
		\hline	
	\end{tabular}
	\caption{Computational costs in CPU time applied to two different, non-redundant systems. 
		\label{tab:comp_costNR}}
\end{table}
\begin{table}
	\begin{tabular}{|l|l|p{3cm}|p{4cm}|}
		\hline	
		Reaction&Architecture&Pre-processing time (s/input)&Average optimization time (s/input/step)\\
		\hline	
		R2&21-30-30$\to$1$\gets$60-42&$6.44\times10^{-3}$&$2.18\times10^{-1}$\\
		\hline	
		Cyclopropylcarbonyl&55-30-30$\to$1$\gets$60-110&$2.20\times10^{-2}$&$6.03$\\
		\hline
	\end{tabular}
	\caption{Computational costs in CPU time applied to two different, redundant systems.
		\label{tab:comp_costR}	}
\end{table}

What is clear in comparing table \ref{tab:comp_costNR} to table \ref{tab:comp_costR} 
is while using the full redundant set of input coordinates reduces the total pre-processing 
time, the per step optimization increases by a factor of at least two for R2 and by a factor 
of approximately ten for the ring-opening of the cyclopropylcarbonyl radical. In the following section, 
one can see from figure \ref{fig:residuals} that at least 5000 steps are required for 
the residual to stabilize, it is thus clear that the time saved in the pre-processing phase by 
using a fully redundant set is overwhelmed by the additional time required in the optimization 
phase, although this would not be the case though for very large systems if the stopping criteria 
mentioned in section \ref{sec:find_nrc} were not enforced.

It is clear then that 
naively applying deep neural networks to large systems is impractical at 
this stage. Importantly, however, the architecture of our system may make 
this issue slightly more tractable. Joining multiple neural networks to 
a single output node might permit the usage of a subset of descriptors 
for each network. Keeping with the example of a molecule of 15 atoms, 
the descriptors for one network might use 14 bond distortions, for 
another 13 bond angles, and yet another 12 dihedral angles. If each 
network contained two hidden layers, only 3289 weights and biases would 
need to be optimised. Dividing the network up in this manner, one 
network for each of the three typical valence descriptors, 
it might 
be possible to extend the applicability of deep neural networks to 
\new{somewhat} larger systems. \new{It is clear, though, that the
  approaches described here are most suitable for highly accurate
  PES-fits for molecules of up to about a dozen atoms. For for the
  description of significantly larger systems, truncation schemes
  like, for example, atomistic neural networks are more promising.}

%%%%%%%%%%%%%%%%%%%%%%%%%%%%%%%%%%%%%%%%%%%%%%%%%%%%%%%%%%%%%%%%%%%%%%%%
\subsection{\sffamily \large The residual and usage of eigenvalues therein}
%%%%%%%%%%%%%%%%%%%%%%%%%%%%%%%%%%%%%%%%%%%%%%%%%%%%%%%%%%%%%%%%%%%%%%%%

\begin{figure}
\begin{center}
  \includegraphics[width=0.9\columnwidth]{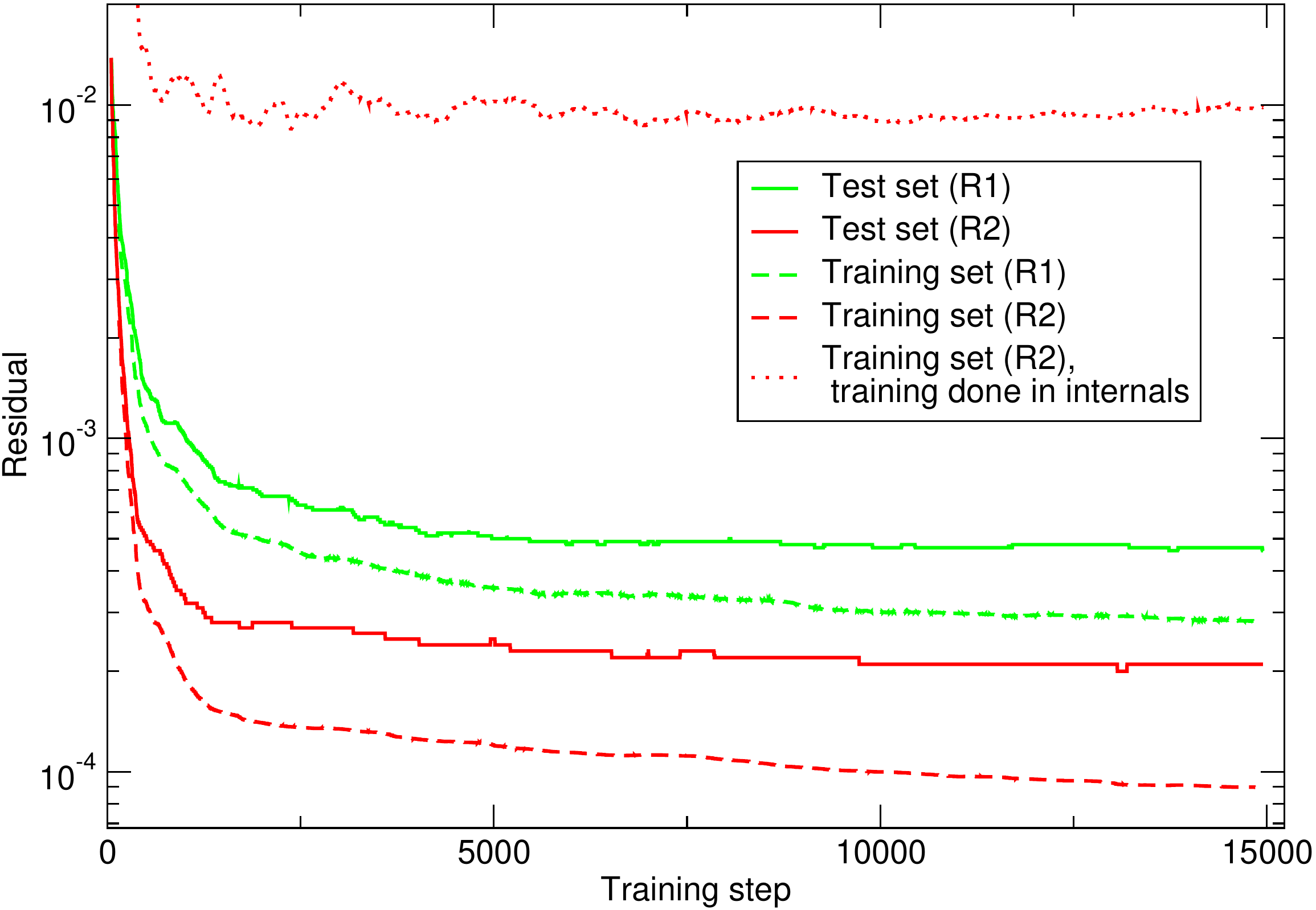}
  \caption{\label{fig:residuals} Running averages of the residuals of
    reactions R1 and R2 during training. Generally, cartesian residuals are
    used for training. For comparison, the dotted line shows the training set
    residual when internal residueals are used for training.}
\end{center}
\end{figure}

In order to give the reader some sense of the learning rate of the neural
network, figure \ref{fig:residuals} shows the reduction of the \emph{test set}
residual in \eqref{eq:residual} with each optimization step. It is noteworthy
that the residual for both reactions stabilizes, given that no regularization
was used as one typically expects the test-set residual to begin increasing at
some point while the training set residual continues to decrease. \new{Figure
  \ref{fig:residuals} also shows the training set residual of a run where the
  error in internal coordinates is minimised (red dotted line). It is clear
  that the result is about two orders of magnitude worse than the minimisation
  of the error in Cartesian coordinates (red dashed line), a result of the 
  fact that the magnitude of the uncertainty introduced by the back-transformation 
  cannot be learned and corrected by the weights and biases in this scheme.}

As can be seen from \eqref{eq:rate}, reaction rate constant calculations depend 
on the eigenvalues of the reactant state and instanton. Accurately 
interpolated Hessians are clearly critical in this regard, yet a 
minimisation procedure that omits eigenvalues from the residual 
optimises all Hessian matrix elements equally. Certain eigenvalues 
are more sensitive to inaccuracies in particular matrix elements than 
others. Clearly then, since the ultimate goal is an accurate 
eigenspectrum for \eqref{eq:rate}, it makes sense to provide the learning procedure 
with eigenvalue residuals in order that the neural network can intuit 
which Hessian elements require greater emphasis on accuracy. This 
extra optimisation option may not always be necessary (as was the case 
with R2), but as we found with R1, it was necessary to set $A_\text{L}=2$,
this had a self-reinforcing, positive influence on the overall residual 
and on the accuracy of the resulting Hessians and eigenvalues, a 
result not otherwise possible.

%%%%%%%%%%%%%%%%%%%%%%%%%%%%%%%%%%%%%%%%%%%%%%%%%%%%%%%%%%%%%%%%%%%%%%%%
\subsection{\sffamily \large  Other coordinate systems}
%%%%%%%%%%%%%%%%%%%%%%%%%%%%%%%%%%%%%%%%%%%%%%%%%%%%%%%%%%%%%%%%%%%%%%%%

It is worth highlighting the fact that while the procedure for 
obtaining non-redundant coordinates outlined in section \ref{sec:find_nrc} 
is appropriate and manageable for the chemical systems treated here, 
the combinatorial scaling of \eqref{eq:combinations}, should one wish 
to test all possible combinations, is prohibitive. We therefore 
acknowledge that the procedure outlined in \cite{Pulay1992,Laude2018} obviates 
this problem, at the expense of complicating somewhat the expressions 
for the coordinate, gradient and Hessian transformations in 
\eqref{eq:transformations} and the subsequent equations for the 
residual.

\begin{figure}
\begin{center}
  \includegraphics[width=0.9\columnwidth]{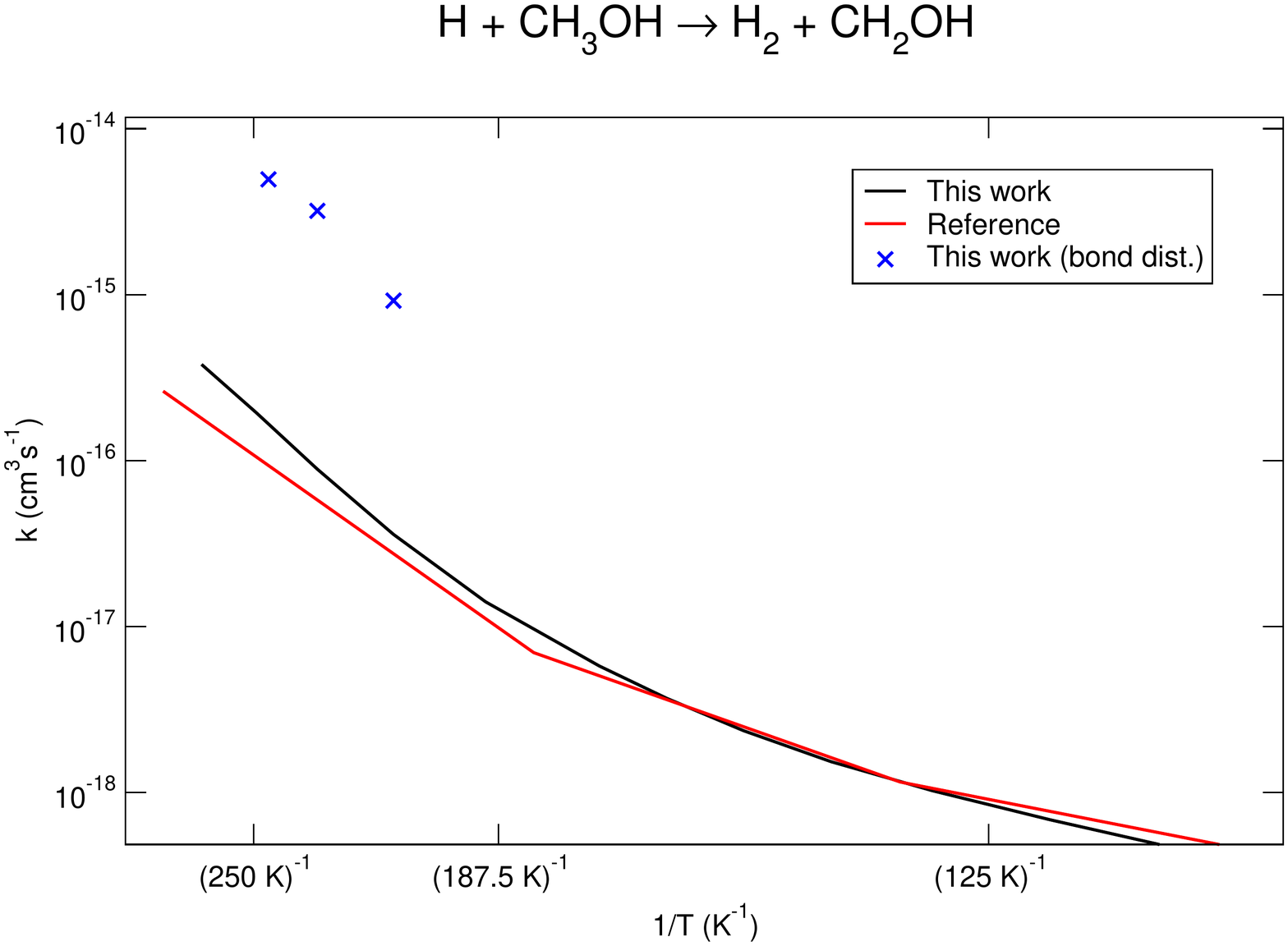}
  \caption{\label{fig:methanol_rate_bdist} Comparision of results using 
  internuclear distances (blue crosses) as descriptors. For this reaction, the 
  rate constants are overestimed and instantons can only be located at higher 
  temperatures.
  }
\end{center}
\end{figure}

Finally, we note that the results presented in section \ref{sec:applications} were produced 
using the inverse internuclear distance coordinate system. The other coordinate 
systems mentioned, internuclear distances and z-matrix coordinates, were also 
tested, yet neither coordinate system was able to yield a residual 
as small as that obtained using inverse internuclear distances. 
As a guide, \figref{fig:methanol_rate_bdist} 
shows the performance of internuclear distances as descriptors applied to R2, 
the rate constants calculated are overestimated by slightly more than one order 
of magnitude for higher temperatures. Moreover, the instanton search 
only reaches convergence at higher temperatures, below 210 K, no 
valid instanton could be found.

%%%%%%%%%%%%%%%%%%%%%%%%%%%%%%%%%%%%%%%%%%%%%%%%%%%%%%%%%%%%%%%%%%%%%%%%
\section{\sffamily \Large Conclusions}
%%%%%%%%%%%%%%%%%%%%%%%%%%%%%%%%%%%%%%%%%%%%%%%%%%%%%%%%%%%%%%%%%%%%%%%%

A neural network PES interpolation scheme has been presented 
capable of calculating energies, gradients and Hessians and which is 
intrinsically independent of the rotation and translation of the molecule 
being modelled. Our design can be readily applied to small molecules 
and is able to produce results which are comparable in 
accuracy to other established methods. The architecture of the 
neural network used here allows for a broad scope of refinements in 
the future with the aim of increasing the range of applicability 
of deep neural networks to reactions involving more than approximately 
ten atoms.

\section{\sffamily \Large Acknowledgements}

This work was financially supported by the European Union's Horizon 2020
research and innovation programme (Grant Agreement Number 646717,
TUNNELCHEM). Computational resources were provided by the state of
Baden-W\"urttemberg through bwHPC and the German Research Foundation (DFG)
through Grant Number INST 40/467-1 FUGG.

%%%%%%%%%%%%%%%%%%%%%%%%%%%%%%%%%%%%%%%%%%%%%%%%%%%%%%%%%%%%%%%%%%%%%%%%
\clearpage
%%%%%%%%%%%%%%%%%%%%%%%%%%%%%%%%%%%%%%%%%%%%%%%%%%%%%%%%%%%%%%%%%%%%%%%%
\bibliography{Q_Chem_Uni_St}
%%%%%%%%%%%%%%%%%%%%%%%%%%%%%%%%%%%%%%%%%%%%%%%%%%%%%%%%%%%%%%%%%%%%%%%%
\clearpage

\end{document}